\documentclass[fleqn,twoside,twocolumn,nofootinbib,showkeys]{revtex4} 
\usepackage[nocpr]{ujp} 

\begin{document}
\title[Physical Nature of Hydrogen Bond]
{PHYSICAL NATURE OF HYDROGEN BOND}%

\author{I.V. Zhyganiuk}
\affiliation{Institute of Environmental Geochemistry, Nat. Acad. of Sci. of Ukraine}
\address{34a, Palladin Ave., Kyiv 03680, Ukraine}
\email{zhyganiuk@gmail.com}
\author{M.P. Malomuzh\,}%
\affiliation{I.I. Mechnikov National University of Odessa}
\address{2, Dvoryans'ka Str., Odessa 65026, Ukraine}

\udk{[539.6:544.353.21:\\[-3pt] 544.142.4]:546.212} 
\razd{\secix}

\autorcol{I.V.\hspace*{0.7mm}Zhyganiuk, M.P.\hspace*{0.7mm}Malomuzh}

\setcounter{page}{960}%

\begin{abstract}
The physical nature and the correct definition of hydrogen bond
(H-bond) are considered.\,\,The influence of H-bonds on the
thermodynamic, kinetic, and spectroscopic properties of water is
analyzed.\,\,The conventional model of H-bonds as sharply directed
and saturated bridges between water molecules is incompatible with
the behavior of the specific volume, evaporation heat, and
self-diffusion and kinematic shear viscosity coefficients of water.
On the other hand, it is shown that the variation of the dipole
moment of a water molecule and the frequency shift of valence
vibrations of a hydroxyl group can be totally explained in the
framework of the electrostatic model of H-bond.\,\,At the same time,
the temperature dependences of the heat capacity of water in the
liquid and vapor states clearly testify to the existence of weak
H-bonds.\,\,The analysis of a water dimer shows that the
contribution of weak H-bonds to its ground state energy is
approximately 4--5 times lower in comparison with the energy of
electrostatic interaction between water molecules.\,\,A conclusion
is made that H-bonds have the same nature in all other cases where
they occur.
\end{abstract}


\maketitle

\begin{flushright}
{\footnotesize\it The hydrogen bond is a phlogiston of the 20-th century.\\
L.~BULAVIN}\vspace*{-4.5mm}
\end{flushright}

\section{Introduction}

During the last 100 years, the concept of hydrogen bond played a large role in
the description of properties of water, alcohols, aqueous alcohol solutions,
and other systems, in which hydrogen bonds were supposed to exist.
Unfortunately, a consistent theory for the latter has not been created within
this time interval.

Hydrogen bonds obviously neither belong to standard chemical bonds
with energies of about $100k_{\mathrm{B}}T_{\mathrm{tr}}$, where
$k_{\mathrm{B}}$ is the Boltzmann constant and $T_{\mathrm{tr}}$ is
the temperature of water's triple point, nor to ionic ones with
energies of the same order of magnitude.\,\,At the same time, the
energies, which are ascribed to hydrogen bonds, exceed those of van
der Waals or dispersion forces by 1--2 orders of magnitude.

M.D.~Sokolov was the first who paid attention to this circumstance
almost 60 years ago \cite{Sokolov}.\,\,He indicated that the energy
of hydrogen bonds approaches the energy of dipole-dipole
interaction.\,\,In such a way, M.D.Sokolov showed that the main
contribution to the hydrogen bond energy is made by the
electrostatic interaction, which does not associated with the
deformation of electron shells in a water molecule.\,\,This idea was
also supported by an outstanding physical chemist Prof.
G.G.~Malenkov during oral discussions on the hydrogen bond problem.
Later, it was shown in works \cite{23Zhyganiuk, Zhyganiuk,
Dolgushin76, Fulton, Barnes, BerendsenVelde} that the deformation of
electron shells in water molecules forming a hydrogen bond is a
small quantity and generates an irreducible contribution to the
energy of interaction between molecules.\,\,It is this contribution
that should be called the hydrogen bond.

In the work by Clementi \cite{Clementi}, the optimum configurations
of hydrogen and oxygen atoms in the isolated water molecule and in
the water molecule located near the Li$^{+}$ ion were
calculated.\,\,The difference between the oxygen--hydrogen distances
in those two configurations amounts to about 0.005~\AA .\,\,This
fact clearly testifies in favor of water molecule models with fixed
invariant positions of model charges, because the distances between
the centers of charges in the water molecule did not change even
under the action of an electric field with a high strength that
occurred near the Li$^{+}$ ion.

The facts indicated above substantiate the application of a number
of model electrostatic potentials proposed or described in works
\cite{Zhyganiuk, Stillinger, Antonchenko, Jorgensen2, Poltev,
Matsuoka, Hirschfelder, Eizenberg, Rieth}.\,\,Those potentials allow
one to explain the thermodynamic and the majority of kinetic
properties of water, as well as many other systems.\,\,At the same
time, the temperature dependence of the heat capacity of water
\cite{Lishchuk} undoubtedly testifies to the necessity to consider
the contributions made by hydrogen bonds (the energy of these bonds
is about $(  1.5$$\div$$3) k_{\mathrm{B}}T_{\mathrm{tr}}$).\,\,A new
view on the character of the intermolecular interaction also follows
from the systematization of liquids proposed by Leonid Bulavin
\cite{Bulavin} (see also works \cite{1Bulavin, 2Bulavin, 3Bulavin}).

In this work, we tried to systematize all main facts that testify to
the necessity of a radical revision of the concept of hydrogen
bonds.\,\,Our conclusions are based on the results obtained for the
temperature dependences of the self-diffusion and kinematic shear
viscosity coefficients, specific volume, evaporation heat
\cite{Fisenko}, and heat capacity of water \cite{Lishchuk,
Lishchuk10, Malomuzh}.\,\,In addition, we analyze the frequency
shift for valence hydrogens in the water molecule \cite{2Zhyganiuk}
and the variation of its dipole moment under the action of
environmental molecules \cite{Zhyganiuk}.\vspace*{-2mm}

\section{Manifestation of Hydrogen\\ Bonds in Transport Processes}

In this section, the manifestations of hydrogen bonds in two most
important kinetic transport phenomena are discussed.\,\,These are
self-diffusion and shear viscosity.\,\,Note that, in order to
exclude the density effects, the behavior of kinematic shear
viscosity will be analyzed.\,\,The main results are obtained with
the use of experimental data measured at the coexistence curve.

From the conventional viewpoint, hydrogen bonds, which couple
neighbor molecules with an energy of about
$10k_{\mathrm{B}}T_{\mathrm{tr}}$, should interfere with the
thermally driven drift of molecules and the relative shift of
adjacent liquid layers.\,\,On the other hand, the rotational motion
of molecules has to be taken into consideration, which follows from
the basic principles of statistical mechanics.\,\,It should be noted
at once that those two statements do not agree with each
other.\,\,If hydrogen bonds are imagined as rods or string segments,
the rotational motion of molecules becomes impossible.\,\,In this
connection, an assumption should be made that the transverse
elasticity of
hydrogen bonds either equals zero or its value does not exceed $k_{\mathrm{B}}%
T_{\mathrm{tr}}$.\,\,The concept of hydrogen bond as a physical
object with so controversial properties is
unsatisfactory.\looseness=-1


\begin{figure}
\vskip1mm
\includegraphics[width=\column]{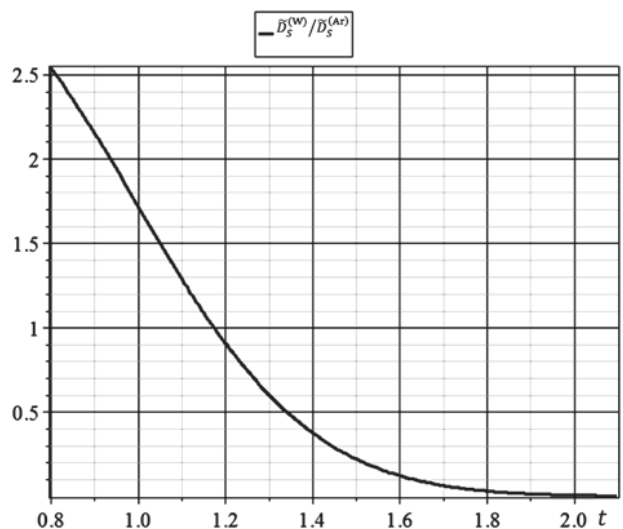}
\vskip-3mm\caption{Dependence of the ratio
$\tilde{\mathrm{D}}_{\mathrm{s}}^{\mathrm{(}w\mathrm{)}}/\tilde{\mathrm{D}}%
_{\mathrm{s}}^{\mathrm{(Ar)}}$ between the diffusion coefficients of
water and argon at the corresponding
temperatures}\label{f21}\vspace*{2mm}
\end{figure}

In order to confirm a conclusion about an unsatisfactory character
of such notions concerning the hydrogen bond, let us consider the
temperature dependences of the relative values of the
self-dif\-fu\-si\-on and viscosity coefficients for water ($w$) and
argon (Ar).\,\,The temperature behaviors of the
self-dif\-fu\-si\-on, $\tilde{D}_{s}^{\mathrm{(}w\mathrm{)}}/\tilde{D}%
_{s}^{\mathrm{(Ar)}}$, and kinematic shear viscosity, ${\tilde{\nu}%
}^{\mathrm{(}w\mathrm{)}}/{\tilde{\nu}^{\mathrm{(Ar)}}}$, coefficient ratios
are depicted in Figs.~1 and 2, respectively (here, ${\tilde{A}}%
=A/A{_{\mathrm{tr}}}$).\,\,At\-ten\-ti\-on should be paid that the
coefficients are compared at the same relative temperatures
$t=T/T_{\mathrm{tr}}^{(i)}$, where $T_{\mathrm{tr}}^{(i)}$ is the
corresponding triple point temperature of the liquid ($i=w$ or Ar).
The states of argon and water in the interval from the triple point
to the critical one and supercooled water are considered.\,\,The
values of self-dif\-fu\-si\-on and viscosity coefficients for water
were taken from the NIST reference database \cite{NIST}, and the
self-dif\-fu\-si\-on coefficients for argon were calculated, by
using the molecular dynamics methods \cite{Naghizadeh, Laghaei}
because of the lack of detailed experimental results.


\begin{figure}
\vskip1mm
\includegraphics[width=\column]{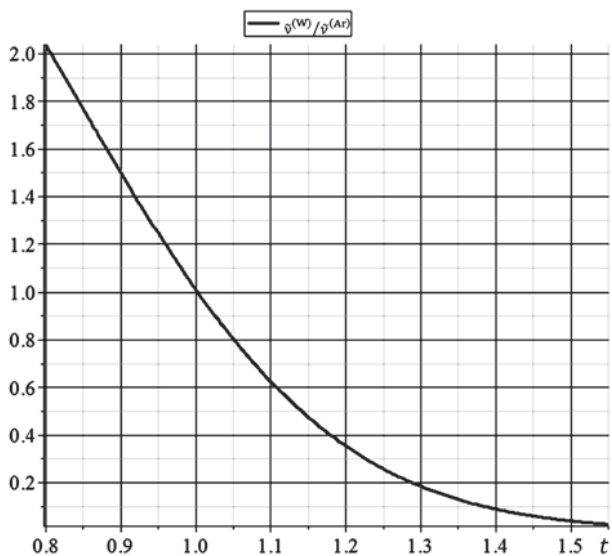}
\vskip-3mm\caption{Dependence of the ratio $\tilde{\nu}_{\mathrm{s}}^{\mathrm{(}w\mathrm{)}%
}/\tilde{\nu}_{\mathrm{s}}^{\mathrm{(Ar)}}$ between the coefficients
of kinematic shear viscosity in water and argon at the corresponding
temperatures  }\label{f22}\vspace*{-2mm}
\end{figure}


\begin{table}[b]
\noindent\caption{Self-diffusion coefficient \\of water molecules in
aqueous solutions \\of single-charged electrolytes {\cite{Douglass}}
}\label{tab1}\vskip3mm\tabcolsep1.8pt
\noindent{\footnotesize\begin{tabular}{|c| c| c| c| c| c| c| c| c|}
 \hline \multicolumn{1}{|c}
{\rule{0pt}{5mm}} & \multicolumn{1}{|c}{}& \multicolumn{1}{|c}{}&
\multicolumn{1}{|c}{$\rm LiBr(30)$}&
\multicolumn{1}{|c|}{$\rm LiI(24.8)$}\\[1.5mm]%
\hline%
\rule{0pt}{5mm}$\mathrm{D}_\mathrm{s}^{(w)}\times10^5, ~\text{сm$^2$/s}$& & & 2.3 & 2.18 \\[1.5mm]%
\hline {\rule{0pt}{5mm}}& \multicolumn{1}{|c} {}&
\multicolumn{1}{|c}{$\rm NaCl(15.9)$}&
 \multicolumn{1}{|c} {$\rm NaBr (16.5)$}& \multicolumn{1}{|c|} {$\rm NaI (16.1)$}\\[1.5mm]%
\hline
\rule{0pt}{5mm}$\mathrm{D}_\mathrm{s}^{(w)}\times 10^5,~\text{сm$^2$/s}$&  & 2.14 & 2.26 & 2.38 \\[1.5mm]%
\hline {\rule{0pt}{5mm}}& \multicolumn{1}{|c} {$\rm KF(15)$}&
\multicolumn{1}{|c}{$\rm KCl(16.1)$}&
 \multicolumn{1}{|c} {$\rm KBr (16.1)$}& \multicolumn{1}{|c|} {$\rm KI (16.4)$}\\[1.5mm]%
\hline
\rule{0pt}{5mm}$\mathrm{D}_\mathrm{s}^{(w)}\times 10^5,~\text{сm$^2$/s}$& 1.99 & 2.44 & 2.68 & 2.8 \\[2mm]%
\hline
\end{tabular}}
\end{table}

As one can see from the dependences exhibited in Figs.~1 and 2,
hydrogen bonds practically do not manifest themselves in the
temperature behavior of the diffusion and viscosity coefficients.
First, the self-diffusion coefficient of water exceeds that of
argon, which completely agrees with a higher mobility of lighter
water molecules (by order of magnitude, the self-diffusion
coefficient ${\tilde{\mathrm{D}}}_{\mathrm{s}}\sim
k_{\mathrm{B}}T/{\nu}$).\,\,Second, the viscosity coefficient for
water is practically by an order of magnitude smaller than that of
argon, which also contradicts the initial statement that hydrogen
bonds may strongly affect the viscosity of water.\,\,Moreover, in
the whole temperature intervals of the liquid state existence, the
self-diffusion and viscosity coefficients of both water and argon
satisfy the relation\vspace*{-2mm}
\begin{equation}
\mathrm{D}_{\mathrm{s}}{\nu}\sim1/{r}_{m},\label{1}%
\end{equation}\vspace*{-5mm}

\noindent where ${r}_{m}$ is the radius of the molecule, i.e., this
combination is close to the corresponding constant for argon or
water.\,\,The invariance of this combination for water testifies
that hydrogen bonds have no substantial relation to the problem of
the water viscosity behavior.

This conclusion evidently correlates with the statement made in work
\cite{Lishchuk10} that the behavior of viscosity is governed by the
averaged interaction potential between molecules.\,\,The averaging
is a consequence of the almost free rotation of water molecules,
which would be impossible in the case of rod-like hydrogen bonds.

Moreover, we attract attention to the absence of a network formed by
hydrogen bonds.\,\,This conclusion immediately results from a
careful analysis of the self-diffusion coefficients and the
mobilities of ions and water molecules in diluted electrolyte
solutions \cite{4Bulavin}.

\section{Mobility of Water\\ Molecules in Electrolyte Solutions}

In this section, the physical nature of the mobility of water
molecules in diluted aqueous solutions of electrolytes, when there
are no more than 15 water molecules per ion, is
discussed.\,\,Attention is focused on the fact that the behavior of
the mobility coefficients of water molecules--or, in other words,
their self-diffusion coefficient--is ultimately determined by the
radii of ionic hard cores.\,\,Hence, the network of hydrogen bonds,
the existence of which is postulated in the overwhelming majority of
works, does not manifest itself.\vspace*{-2mm}

\subsection{Self-diffusion coefficients\\ of water molecules}

First of all, we would like to attract attention to how the behavior
of the self-diffusion coefficients of water molecules depends on the
dimensions of cations and anions.\,\,The corresponding values of
self-diffusion coefficients $\mathrm{D}_{\mathrm{s}}^{(w)}$ of water
molecules in several diluted electrolyte solutions at the
temperature $T=296$~K are listed in Table~1.\,\,In each of three
table rows, the cation remains the same, i.e., Table~1 exhibits the
dependence of $\mathrm{D}_{\mathrm{s}}^{(w)}$ on the anion
size.\,\,The solution concentration in the table is presented by the
number $z_{w}$ of water molecules per ion.\,\,The $z_{w}$-value is
shown in parentheses near the chemical formula of electrolyte, e.g.,
NaCl(15,9).\,\,Hence, the minimum value $z_{w}=15$ corresponds to an
electrolyte concentration of 3.3~mol.\%.\,\,At such a concentration,
the mutual influence of cations and anions can be neglected with a
satisfactory accuracy.\,\,Table~1 should also include the
self-diffusion coefficient of water molecules in the
solution CsI(17,4), namely, $\mathrm{D}_{\mathrm{s}}^{(w)}=2{.}71\times10^{5}~\mathrm{cm}%
^{\mathrm{2}}\mathrm{/s}$.\,\,In accordance with works
\cite{Blanckenhagen, LokotoshJ.Chem.Eng.Data}, the self-diffusion
coefficient of molecules in water at the same temperature $T=296$~K
equals\vspace*{-1mm}
\begin{equation}
\mathrm{D}_{\mathrm{s}}^{(w)}=2{.}35\times10^{-5}~\mathrm{cm}^{\mathrm{2}}\mathrm{/s}.\label{Ds}%
\end{equation}\vspace*{-5mm}

One can see that the self-diffusion coefficients of water molecules
grow as the sizes of cations (the data in columns) and anions (the
data in rows) increase, which could testify to a destruction of the
hydrogen bond network. At the same time, in this case, water
molecules would have freely surrounded electrolyte ions to form
hydration spheres and to give rise to a substantial reduction of the
self-diffusion coefficients of water molecules, which contradicts
the data quoted in Table~1.\vspace*{-2mm}

\subsection{Features in the dependence\\ of the self-diffusion coefficients of
water\\ molecules on the ion size}

To make the analysis of the results quoted in Table~1 more
comprehensive, let us consider the correlations between the
self-diffusion coefficients of water molecules in electrolyte
solutions and the radii of dissolved ions.\,\,The considered radii
of ions (i)~were determined from crystallographic data, (ii)~were
selected in a way to favor the correct reconstruction of the
molecular dynamics in electrolyte solutions with the use of computer
simulation methods, and (iii)~were evaluated from the ionic
polarizability \mbox{values.}

The first row in Table~2 contains the values of crystallographic
radius $r_{c}$ of ions \cite{Nightingale}.\,\,The second row
contains the radii of ions $r_{\sigma}$ determined in computer
experiments aimed at the description of the dispersion (van der
Waals) interaction between ions and water molecules \cite{Koneshan}.
The radii of ions $r_{\alpha}$, which were determined from ionic
polarizabilities $\alpha$, by using the formula
\cite{House}\vspace*{-1mm}
\begin{equation}
r_{\alpha}=1.5\alpha^{1/3},\label{pol}%
\end{equation}
are quoted in the third row.\,\,The radii $r_{c}$, $r_{\sigma}$, and
$r_{\alpha}$ will be referred to as hard ionic radii.

From Tables~1 and 2, it follows that the dependences of the self-diffusion
coefficients of water molecules on the radii of ionic hard cores demonstrate
the following regularities:

1) for diluted lithium and sodium electrolyte solutions, in which
$r_{c}<\frac{1}{2}l_{w}$, where $l_{w}$ is the average distance
between the
oxygen atoms in neighbor water molecules, the inequality $\mathrm{D}_{\mathrm{s}}^{(w)}%
(\mathrm{el})<\mathrm{D}_{\mathrm{s}}^{(w)}$ is obeyed.\,\,The
inequality $r_{c}<\frac{1}{2}l_{w}$ is obviously violated only for
Cs$^{+}$.\,\,In the solutions of potassium electrolytes, in which
$r_{c}(\mathrm{K^{+}})\sim\frac{1}{2}l_{w}$, a transition from the
previous inequality between the self-diffusion
coefficients of water molecules to the inequality $\mathrm{D}_{\mathrm{s}}^{(w)}(\mathrm{el}%
)>$ $>\mathrm{D}_{\mathrm{s}}^{(w)}$ is observed;

2) in electrolyte solutions with a fixed cation, but for the lithium
one, the self-diffusion coefficients of water molecules grow
together with the anion radius;

3) in the lithium electrolytes, the character of the dependence of
$\mathrm{D}_{\mathrm{s}}^{(w)}(\mathrm{el})$ on the anion radius is
opposite to that described in the previous item.


\begin{table}[b]
\noindent\caption{Hard core radii (three upper\\ rows) and Stokes
radii of cations and anions}\label{tab2}\vskip3mm\tabcolsep4.4pt

\noindent{\footnotesize\begin{tabular}{|l|c|c|c|c|c|c|c|c|}
 \hline \multicolumn{1}{|c|}
{\rule{0pt}{5mm}}& \multicolumn{1}{|c|}{$\rm Li^{+}$} &
\multicolumn{1}{|c|}{$\rm Na^{+}$} & \multicolumn{1}{|c|}{$\rm
K^{+}$}& \multicolumn{1}{|c}{$\rm Cs^{+}$ }&
\multicolumn{1}{|c|}{$\rm F^{-}$}& \multicolumn{1}{|c|}{$\rm
Cl^{-}$} & \multicolumn{1}{|c|}{$\rm Br^{-} $} &
\multicolumn{1}{|c|}{$\rm I^{-}$}\\[2mm]%
\hline%
 \rule{0pt}{5mm}$r_c$,~{\AA} &0.6~\,& 0.95& 1.33&1.69  & 1.36& 1.81 & 1.95 & 2.16 \\%
$r_{\sigma}$,~{\AA} &0.76& 1.3~\,& 1.67& 1.94& 1.56& 2.2~\, & 2.27 & 2.59 \\%
$r_{\alpha}$,~{\AA} &0.45& 1.12& 1.41& 2.02& 1.51& 2.33 & 2.55 & 2.93 \\%
$r_s^{(\mu)}$,~{\AA} &2.38& 1.84& 1.25&1.19 & 1.66& 1.21 & 1.18 & 1.19 \\%
$r_c^{(D)}$,~{\AA} &1.91& 1.88& 1.14&1.15 & 1.77& 1.33 & 1.22 & 1.35 \\[2mm]%
\hline
\end{tabular}}
\end{table}

Since the concentrations of different single-charged cations and anions are
close to one another, the differences in the behavior of $\mathrm{D}_{\mathrm{s}}^{(w)}%
(\mathrm{el})$ can be associated with geometrical factors and their
different action on the structure of local environment (hydration
effects).\,\,The former possibility should be rejected, because the
geometrical obstacles should diminish with the reduction of the
cation radius, which contradicts experimental data.\,\,At the same
time, a local restructuring of water in close vicinities of cations
and anions turns out more appreciable for larger radii of their hard
cores.

It should be noted that the effect of local restructuring in the
solution is insignificant, because an increase or a reduction of the
self-diffusion coefficient for water molecules in most cases does
not exceed 10\% and is proportional to the molar concentration of
electrolyte admixtures.\,\,All those facts are completely
incompatible with the statement about the presence of a developed
network of hydrogen bonds in water and diluted aqueous electrolyte
solutions.

\subsection{Nonideality degree of electrolyte solutions}

The results obtained should be appended with the data concerning the
nonideality degree of diluted electrolyte solutions, which were
obtained in work \cite{4Bulavin}.\,\,According to the cited work,
the nonideality degree can be calculated using the relation
\begin{equation}\label{eq514}
\delta  = \frac{\rho }{{\rho _w^{(0)} }}\left[ {1 + x\left(\!
{\frac{{\rho _w^{(0)} }}{{\rho _{\mathrm{el}}^{(0)} }} - 1}
\!\right)}\! \right]
 - 1 ,
\end{equation}
where ${{\rho_{w}^{(0)}}}$ and ${\rho_{\mathrm{el}}^{(0)}}$ are the
densities of water and electrolyte, respectively, in the liquid or
amorphous state, and $\rho$ is the density of the aqueous
electrolyte solution.\,\,The densities of some electrolyte solutions
can be found in Table~3, and the degrees of their nonideality,
$\delta$, are given in Table~4.


\begin{table}[b!]
\noindent\caption{Densities of aqueous electrolyte\\ solutions at
the fixed concentration
\boldmath$x_{\mathrm{el}}=4\mathrm{~wt.\%}$}\label{tab3}
\vskip3mm\tabcolsep13.6pt

\noindent{\footnotesize\begin{tabular}{|c|c|c|c|c|}
 \hline \multicolumn{1}{|c|}
{\rule{0pt}{5mm}} & \multicolumn{1}{|c|}{$\mathrm{F}^{-}$}&
\multicolumn{1}{|c|}{$\mathrm{Cl}^{-}$}& \multicolumn{1}{|c|}{$
\mathrm{Br}^{-}$}&
\multicolumn{1}{|c|}{$\mathrm{I}^{-}$}\\[2mm]%
\hline%
\rule{0pt}{5mm}$\mathrm{Li}^{+}$\,&         & 1.02~\,  & 1.031 & 1.033 \\%
$\mathrm{Na}^{+}$&         & 1.026 &         & 1.035 \\%
$\mathrm{K}^{+}$\,\, & 1.032 &         & 1.029 & 1.032 \\%
$\mathrm{Cs}^{+}$& 1.034 & 1.035 &         &         \\[2mm]%
\hline
\end{tabular}}
\end{table}

\begin{table}[b!]
\vskip3mm \noindent\caption{Parameters of solution\\ nonideality at
the fixed concentration
\boldmath$x_{\mathrm{el}}=4\mathrm{~wt.\%}$}\label{tab4}
\vskip3mm\tabcolsep18.0pt

\noindent{\footnotesize\begin{tabular}{|c|c|c|c|}
 \hline
\multicolumn{1}{|c|} {\rule{0pt}{5mm}} &
\multicolumn{1}{|c|}{$\mathrm{Cl}^{-}$}& \multicolumn{1}{|c|}{$
\mathrm{Br}^{-}$}&
\multicolumn{1}{|c|}{$\mathrm{I}^{-}$}\\[2mm]%
\hline%
\rule{0pt}{5mm}$\mathrm{Li}^{+}$\,&  $-$0.001~\,\, &         &  \\%
$\mathrm{Na}^{+}$&  0.003   &         & \\%
$\mathrm{K}^{+}$\,\, &            & 0.003 & 0.004 \\%
$\mathrm{Cs}^{+}$&  0.005   &         & \\[2mm]%
\hline
\end{tabular}}
\end{table}

Note that the values of nonideality parameter obtained by formula
(\ref{eq514}) are averaged over the number of water molecules that
surround the corresponding ion.\,\,One can see that relatively small
values of the mobility of lithium cations and the self-diffusion
coefficient of water molecules in the lithium electrolytes
correspond to negative nonideality degrees.\,\,This fact testifies
that lithium cations do not promote the formation of hydration
spheres around them with the density exceeding that of water.\,\,In
other cases, if one may talk about the formation of hydration
spheres, the smallness of the parameter $\delta$ testifies that
their influence on the density in diluted electrolyte solutions is
weak.

From the facts mentioned above, it follows that

(i)~the key role in governing the transport properties of aqueous
electrolyte solutions--first of all, the behavior of the mobility
coefficients of ions and water molecules--is played by hard cores of
those objects;

(ii)~the conventional scenario that ions in aqueous solutions move
through \textquotedblleft voids\textquotedblright\ existing in the
hydrogen bond network is incorrect;

(iii)~hydrogen bonds between water molecules have to be considered
as a convenient model, which to a certain extent reflects the
existence of correlations between the dipole moments of molecules,
as well as multipole moments of higher orders; and

(iv)~the role of hydration effects is insignificant and can be taken
into account in the framework of the thermodynamic perturbation
theory.

\section{Experimental Evidence\\ for the Similarity of the
Thermodynamic\\
Properties of Water and Argon}

In this section, some facts testifying to the thermodynamic
similarity between water and argon and, hence, calling into question
the conventional viewpoint about the crucial role of hydrogen bonds
in the formation of water properties are presented.\,\,With that end
in view, let us consider the temperature dependences for the
simplest quantities: the fractional volume (this is one of the
mechanical characteristics of the system) and the evaporation heat
(the most important among the thermal parameters).\,\,In parallel,
the behavior of ordinary water and its heavy counterpart, which
considerably differ from each other by the character of molecular
rotational motion, will be analyzed.

The temperature dependences of the fractional volumes
$\upsilon^{(i)}$ of water ($i=w$) and argon ($i=\mathrm{Ar}$) at
their coexistence curves will be compared in the spirit of the
similarity principle for the corresponding states of the system
\cite{Hilbert}.\,\,This means that the ratio ${R_{\upsilon
}^{(\mathrm{H}_{2}\mathrm{O})}(t)=\tilde{\upsilon}^{(\mathrm{H}_{2}%
\mathrm{O})}(t)/\tilde{\upsilon}^{(\mathrm{Ar})}(t)}$ between the
normalized volumes
${\tilde{\upsilon}^{(i)}(t)=\upsilon^{(i)}(t)/\upsilon_{c}^{(i)}}$,
where ${\upsilon_{c}^{(i)}}$ means the $\upsilon^{(i)}(t)$-value at
the corresponding critical point, should be considered as a function
of the dimensionless temperature $t=T/T_{c}^{(i)}$, where
$t=T/T_{c}^{(i)}$ is the critical temperature of the $i$-th
liquid.\,\,As is seen from Fig.~3, practically in the whole
temperature interval of water existence in the liquid state,
$0.42<t<0.9$, the temperature dependences of the fractional volumes
of water
and argon are similar.\,\,Their ratio $R_{\upsilon}^{(\mathrm{H}_{2}\mathrm{O}%
)}(t)$ in the temperature interval $0.55<t<0.8$ is approximated with a
satisfactory accuracy by the linear dependence
\begin{equation}
R_{\upsilon}^{(\mathrm{H}_{2}\mathrm{O})}(t)=r_{0}^{(\upsilon)}+4r_{\mathrm{H}%
}^{(\upsilon)}(1-\lambda_{\upsilon}\,t)\label{Rupsilon}%
\end{equation}
with the coefficients
\begin{equation}\label{r0H2O}
\begin{array}{l}
 \displaystyle\mathrm{H}_2 \mathrm{O} : r_0^{(\upsilon )}  = 0.827~r_\mathrm{H}^{(\upsilon )}
  = 0.021~ \lambda _\upsilon=
 0.83,\\[3mm]
  \displaystyle\mathrm{D}_2 \mathrm{O} : r_0^{(\upsilon )}  = 0.851~ r_\mathrm{H}^{(\upsilon )}
   = 0.021~\lambda _\upsilon=
 0.83,
 \end{array}
\end{equation}
i.e.\,\,it remains almost constant.

The deviations of the ratio $R_{\upsilon}(t)=\upsilon^{(\mathrm{D}%
_{2}\mathrm{O})}(t)/$ $/\upsilon^{(\mathrm{H}_{2}\mathrm{O})}(t)$
from unity is appreciable only for at $t\ll$ $\ll0.55$ and in a
vicinity of the critical point. However, they do not exceed 4\%.

A comparison between the evaporation heats for water and argon is even more
intriguing (see Fig.~4).\,\,The deviation of $R_{q}^{(\mathrm{H}_{2}\mathrm{O}%
)}(t)$ from an approximate value of about 6.2 does not exceed 1.1\%.

Analogously to formula (\ref{Rupsilon}), the ratio $R_{q}^{(\mathrm{H}%
_{2}\mathrm{O})}(t)$ in the temperature interval $0.55<t<0.8$ is quasilinear,
\begin{equation}
R_{q}^{(\mathrm{H}_{2}\mathrm{O})}(t)=r_{0}^{(q)}+4r_{\mathrm{H}}%
^{(q)}(1-\lambda_{q}\,t),\label{Rqw}%
\end{equation}
where%
\begin{equation}\label{r0qw}
\begin{array}{l}
 \displaystyle
 \mathrm{H}_2 \mathrm{O} :  r_0^{(q)}  = 6.137~~ r_\mathrm{H}^{(q)}
  = 0.080 ~  \lambda _q   = 0.85
 \\[3mm]
 \displaystyle \mathrm{D}_2 \mathrm{O} : r_0^{(q)}  = 6.084 ~ r_\mathrm{H}^{(q)}
   = 0.182~~\lambda _q   =
 0.85.
 \end{array}
\end{equation}
Moreover, the linear functions $1-\lambda_{\upsilon}t$ and
$1-\lambda_{q}t$ in Eqs.~(\ref{Rupsilon}) and (\ref{Rqw}),
respectively, are almost identical, which testifies to their common
origin.\,\,In works \cite{Lishchuk, Lishchuk10}, it was shown that
those functions are generated by weak hydrogen bonds.

Small deviations of ${R_{\upsilon}^{(i)}(t)}$ and ${R_{q}^{(i)}(t)}$
from their constant values evidently testify to a weak effect of
hydrogen bonds and a similarity of intermolecular potentials in
water and argon.\,\,The latter conclusion has a completely natural
explanation.\,\,The behavior of the fractional volume and the
evaporation heat for water is governed by the averaged potential of
interaction between molecules.\,\,In its turn, the potential
averaging is a direct consequence of the rotational motion of water
molecules.\,\,As a result, the anisotropic effects associated with
weak hydrogen bonds become practically smoothed out.


\begin{figure}
\vskip1mm
\includegraphics[width=\column]{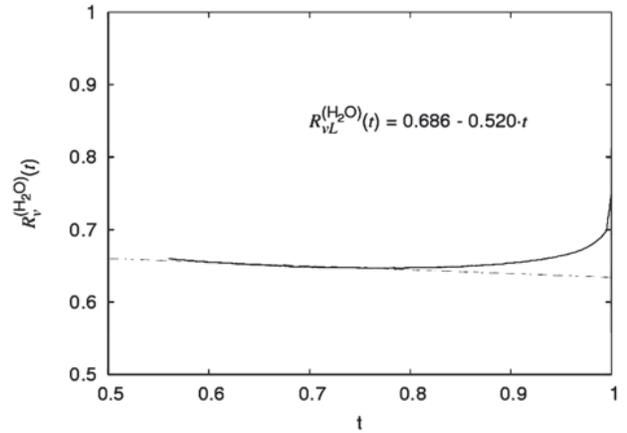}
\vskip-3mm\caption{Temperature dependences of the ratio {$R_{\upsilon}^{(\mathrm{H}%
_{2}\mathrm{O})}(t)=$ $=\tilde{\upsilon}^{(\mathrm{H}_{2}\mathrm{O})}%
(t)/\tilde{\upsilon}^{(\mathrm{Ar})}(t)$} calculated at the
coexistence curves of water and argon in accordance with
experimental data {\cite{NIST}}
 }\label{fRnuWt}
\end{figure}

\begin{figure}
\vskip1mm
\includegraphics[width=\column]{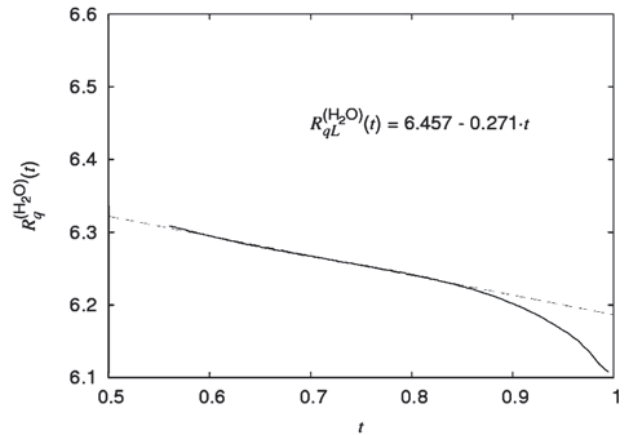}
\vskip-3mm\caption{Temperature dependences of the ratio
$R_{q}^{(\mathrm{H} _{2}\mathrm{O})}(t)=$
$=q^{(\mathrm{H}_{2}\mathrm{O})}(t)/q^{(\mathrm{Ar})}(t)$ in
accordance with experimental data {\cite{NIST}}
 }\label{fRq0Wt}
\end{figure}

\section{Vibration Frequencies\\ of Hydrogen Atoms in the Water Molecule\\ in
Vapor, Water, and Ice}

Let us discuss the frequency shift for the longitudinal (valence)
vibrations of the hydrogen ion H$_{1}^{+}$ in a water molecule,
which lies close to the line connecting the centers of mass of
oxygen atoms in two neighbor water molecules composing a dimer (see
Fig.~5).\,\,The corresponding vibration frequencies are determined
by the formula
\begin{equation}\label{eq14}
\omega _{\parallel}  \approx \sqrt {\frac{{{\rm K}_{{\rm
{rr}}}^{{\rm (1)}} }}{{{\rm M}_{{\rm Re}} }}} ,
\end{equation}
where the force constant (the dimensionless quantities $\tilde{\Phi}%
=\Phi/k_{\mathrm{B}}T_{\mathrm{tr}}$ and $\tilde{r}_{{{{\mathrm{H}}}}_{k}%
}=r_{{{{\mathrm{H}}}}_{k}}/\sigma$, where $\sigma=298$~\AA \ is the oxygen
diameter, will be used) is defined in a standard way:
\begin{equation}\label{eq15}
{\tilde {\rm K}_{\rm {rr}}^{{\rm (1)}}  = \frac{{\partial ^{\rm 2}
{\rm \tilde \Phi }_{\rm 2}^{{\rm (1)}} }} {{\partial {\rm \tilde
r}_{{\rm H}_{\rm 1} }^{\rm 2} }}} |_ {{\rm \tilde r'}_{{\rm H}_{\rm
1} } },
\end{equation}


\begin{figure}
\vskip1mm
\includegraphics[width=7cm]{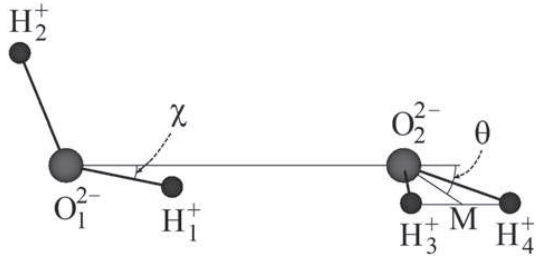}
\vskip-3mm\caption{Configuration of a linear dimer of water
molecules. The equilibrium angle values are {$\chi=3.76^{\circ}$ and
$\theta=41.1^{\circ}$}}\label{f2}
\end{figure}
\begin{figure}
\vskip3mm
\includegraphics[width=\column]{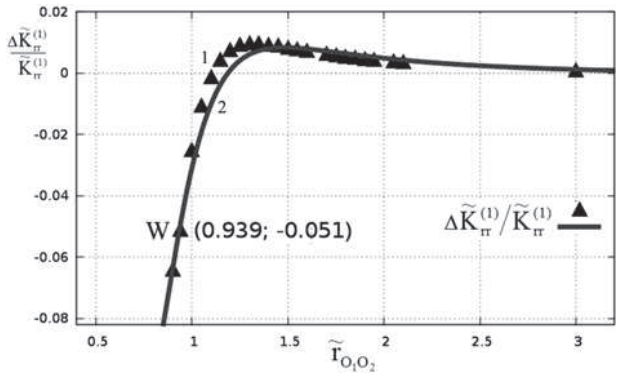}
\vskip-3mm\caption{Dependences of the relative change $\Delta\tilde{\mathrm{K}}_{\mathrm{rr}}^{{(1)}%
}/\tilde{\mathrm{K}}_{\mathrm{rr}}^{{(1)}}$ of the force constant on
the distance $\tilde
{r}_{{{\mathrm{O}}_{\mathrm{1}}{\mathrm{O}}_{\mathrm{2}}}}$ in the
standard dimer: with regard for a mutual tuning of molecular
orientations (\textit{1}, triangles) and for the fixed orientation
of molecules as in the configuration in Fig.~5 (\textit{2},
curve)}\label{f_3}
\end{figure}

\begin{table}[b]\label{tab5}
\noindent\caption{Force constants for water
molecules}\vskip3mm\tabcolsep14.3pt

\noindent{\footnotesize\begin{tabular}{|l|c|c|c|}
\hline%
 \multicolumn{1}{|c|}{\rule{0pt}{5mm}}%
 & \multicolumn{1}{|c|}{$\frac{{\partial ^2 \tilde \Phi }}{{\partial \tilde r_1^2 }}$ }
 & \multicolumn{1}{|c|}{$\frac{{\partial ^2 \tilde \Phi }}{{\partial \theta ^2 }}$}
 & \multicolumn{1}{|c|}{$\frac{{\partial ^2 \tilde \Phi }}{{\partial \tilde r_1 \partial \theta }}$}
  \\[2mm]
  \hline
 \rule{0pt}{5mm}GSD  & 256.98 & 182.15 & 33.439 \\
   Experim.~\cite{Stillinger} & 256.98 & 190.64 & 33.439 \\ [2mm]%
  \hline
\end{tabular}}
\end{table}

\noindent and the reduced mass of the oxygen--hydrogen system in the
water molecule approximately equals
\begin{equation}\label{eq16}
{\rm M}_{{\rm Re}}  \approx \frac{{\left( {{\rm M}_{\rm O} + {\rm
M}_{\rm H} } \right){\rm M}_{ \rm H} }} {{\left( {{\rm M}_{ \rm O} +
{\rm M}_{ \rm H} } \right)
 + {\rm M}_{\rm H} }}.
\end{equation}

The approximate character of Eq.~(\ref{eq16}) is explained by the
fact that hydrogen H$_{2}^{+}$ (see Fig.~5) is not located at the
line connecting the centers of mass of oxygen atoms.\,\,The double
derivative in Eq.~(\ref{eq15}) is assumed to be taken at the point
$\tilde{r}_{\mathrm{H}_{\mathrm{1}}}^{\prime }$ determined by the
equation
\begin{equation}\label{eq17}
 {\frac{{\partial \left( {\tilde \Phi _{{{\mathrm
{1}}}}^{{{\mathrm {(1)}}}}
 (  {\widetilde {\mathrm {\bf  {r} }}}_{{{{\mathrm H}}}_{{{\mathrm 1}}} } ,
  {\widetilde {\mathrm {\bf  {r} }}}_{{{{\mathrm H}}}_{{{\mathrm 2}}} } )
  \! + \!\tilde \Phi _{{{\mathrm {Int}}}} ({\bf \tilde r}_{{{\rm O}_{\rm 1} {\rm H}_{\rm 3}} }
,{\bf \tilde r}_{{{\rm O}_{\rm 1} {\rm H}_{\rm 4}} },
   {\bf \tilde r}_{{ {\mathrm O}_{\mathrm 1}
{\mathrm O}_{\mathrm 2} }} )} \right)}}{{\partial {\rm \tilde
r}_{{{\rm H}_{\rm 1} }} }}} \biggl|_{{\rm \tilde r'}_{{{\rm H}_{\rm
1} }} }
 \!\!= 0.
\end{equation}

It should be noted that Eq.~(\ref{eq17}) gives rise to only an
insignificant displacement
$\Delta\mathrm{\tilde{r}}_{{\mathrm{H}_{\mathrm{1}}}}$ of the
equilibrium position of hydrogen H$_{1}^{+}$.\,\,Assuming that
\[
{\rm \tilde r}_{{\mathrm{H}_{\mathrm{1}}}}^{\prime}={\rm \tilde r}_{{\mathrm{H}%
_{\mathrm{1}}}}+\Delta{\rm \tilde r}_{{\mathrm{H}_{\mathrm{1}}}},
\]
where
\[
|\Delta{\rm \tilde r}_{{\mathrm{H}_{\mathrm{1}}}}|\ll{\rm \tilde r}_{{\mathrm{H}%
_{\mathrm{1}}}},
\]
the value of $\Delta{\rm \tilde r}_{{\mathrm{H}_{\mathrm{1}}}}$ can
be determined with the help of a more simple equation,
\begin{equation}\label{eq18}
{\widetilde {\mathrm K}_{{\mathrm {rr}}}^{{\mathrm{(1)}}} \Delta
{\rm \tilde r}_{{\rm H}_{\rm 1} }  +  \vec \nabla _{{\rm \tilde
r}_{{\rm H}_{\rm 1} } } \widetilde \Phi _{{\rm  {Int}}} ({ \tilde  {
\rm {\bf r}}}_{ {\rm O}_{\rm 1} {\rm H}_{\rm 3} } ,{ \tilde  { \rm
{\bf r}}}_{\scriptscriptstyle {\rm O}_{\rm 1} {\rm H}_{\rm 4} }, {
\tilde { \rm {\bf r}}}_{{\rm O}_{\rm 1} {\rm O}_{\rm 2} }
)}\biggl|_{{\rm \tilde r}_{{{\mathrm H}_{\mathrm 1}} }  =
   {\rm \tilde r'}_{{\rm H}_{\rm 1} }}  \!\!= 0 ,
\end{equation}
where ${\tilde{\mathrm{K}}}_{\mathrm{rr}}^{{\mathrm{(1)}}}$ is the
elasticity coefficient for the bond between the hydrogen and oxygen
atoms in the monomer.

One may get convinced (see work \cite{2Zhyganiuk}) that, by the
order of magnitude in the framework of the electrostatic model, the
ratio $\Delta{\rm \tilde r}_{{\mathrm{H}_{\mathrm{1}}}}/{\rm \tilde
r}_{{\mathrm{H}_{\mathrm{1}}}}$, where
$\Delta{\rm \tilde r}_{{\mathrm{H}_{\mathrm{1}}}}=$ $={\rm \tilde r}_{{\mathrm{H}%
_{\mathrm{1}}}}^{\prime}-{\rm \tilde
r}_{{\mathrm{H}_{\mathrm{1}}}}$, satisfies the inequality
\[
\Delta {\rm \tilde r}_{{{\rm H}_{\rm 1}} } / {\rm \tilde r}_{{{\rm
H}_{\rm 1}} } \sim ({ {\rm \tilde r}_{ {{\rm H}_{\rm 1}} } / {{\rm
\tilde r}_{{{\mathrm O}_{\mathrm 1} {\mathrm O}_{\mathrm 2}} } } }
)^{4}  \leq 0.02.
\]
Therefore, the displacement of a hydrogen atom can be neglected
practically for all distances between the oxygen atoms in the water
dimer.

The values of force constants for the water molecule, which
correspond to the GSD potential, are listed in Table~5.\,\,For the
sake of comparison, Table~5 also contains the corresponding force
constants determined experimentally.

The force constant for symmetric valence vibrations in the
equilibrium configuration of a dimer shown in Fig.~5 can be
calculated by formula (\ref{eq15}). The dependence of the relative
change of a force constant on the distance between the oxygens in
the dimer is depicted in Fig.~6.\,\,Point~W with
the coordinates $(0.939,-0.051)$ corresponds to the distance $r{_{{{\mathrm{O}%
}_{\mathrm{1}}{\mathrm{O}}_{\mathrm{2}}}}}=2{.}8$~{\AA} between the
oxygen atoms, which is typical of water near its triple
point.\,\,The relative change of the constant of symmetric valence
vibrations at this point equals
$\Delta\tilde{\mathrm{K}}_{\mathrm{rr}}^{{(1)}}/\tilde{\mathrm{K}}_{\mathrm{rr}}^{{(1)}}=-0.05$.
Hence, the values of the force constant of valence vibrations at the
distances between the oxygen atoms corresponding to a dimer at
equilibrium and to liquid water differ from each other by 5\%.

\subsection{Results of calculations\\ of the valence vibration frequencies\\ for
hydrogen atoms in the water molecule}

The frequency shift of hydrogen valence vibrations in the water
molecule depends on the water phase state and can reach several
hundreds of inverse centimeters (Table~6).\,\,In this work, we
assume that the main contribution to the experimentally observed
frequency shift is made by electrostatic forces associated with
multipole moments of water molecules.

The major result of our research consists in that the electrostatic
forces really induce the frequency shifts, which agree with
experimental data by both the shift direction and the order of
magnitude.\,\,The frequency shift of valence vibrations equals
\[
{\rm \Delta \omega } \approx \frac{{\rm 1}}{{\rm 2}}{\rm \omega
}_{\rm 0} \frac{{\Delta \widetilde {\rm K}_{ \rm {rr}}^{{(1)}}
}}{{\widetilde {\rm K}_{ \rm {rr}}^{ {(1)}} }},
\]
where $\omega_{0}\approx3657~$cm$^{-1}$ is the vibration frequency
for an isolated water molecule.\,\,The relative increase of the
elastic constant
$\Delta\tilde{\mathrm{K}}_{\mathrm{rr}}^{{(1)}}/\tilde{\mathrm{K}}_{\mathrm{rr}}^{{(1)}}$
 at $r_{{{\mathrm{O}%
}_{\mathrm{1}}{\mathrm{O}}_{\mathrm{2}}}}=2.8$~{\AA} amounts to
$-0.05$, i.e.\,\,$\Delta\omega\approx-91.43$~\textrm{cm}$^{-1}$.
Hence, the frequency shift sign for a dimer correlates with those in
liquid water and ice.\,\,The shear moduli are identical by order of
magnitude, but, nevertheless, they are considerably
different.\,\,This fact has a simple qualitative
interpretation.\,\,The total electric field that acts on a water
molecule in the liquid is, on the average, a little larger than that
acting from the neighbor molecule in the dimer.\,\,An insignificant
increase of the electric field strength in the liquid is connected
with a weakly ordered arrangement of the centers of mass of
molecules and the orientations of its nearest neighbors.\,\,As a
consequence, owing to the superposition principle, only a weak
strengthening of the electric field in the molecule occurs.\,\,The
opposite situation takes place in ice.

Let us discuss the change of elastic constant in the standard dimers (see
Fig.~5) at $r_{{{\mathrm{O}}_{\mathrm{1}}{\mathrm{O}}_{\mathrm{2}}}%
}=2.85$~{\AA}.\,\,This is a distance between the oxygen atoms of
water molecules in an argon matrix.\,\,According to our
calculations, the relative increase of the elastic constant
$\Delta\tilde{\mathrm{K}}_{\mathrm{rr}}^{{(1)}}/\tilde
{\mathrm{K}}_{\mathrm{rr}}^{{(1)}}$ for this dimer configuration
amounts to $-0.0443$ for the fixed orientation of molecules as in
the standard dimer and to $-0.0432$ if the orientations of molecules
can be adjusted.\,\,These variations of the elastic
constant correspond to a valence vibration frequency of 3576~\textrm{cm}%
$^{-1}$ in the former case and 3578~\textrm{cm}$^{-1}$ in the latter
one.\,\,It should be noted that the relatively small value of
orientational contribution is explained by the fact that the
parameter of the system
$r_{{{\mathrm{O}}_{\mathrm{1}}{\mathrm{O}}_{\mathrm{2}}}}%
=2.85$~{\AA} is in the interval, where the dependence of the
repulsion energy between molecules monotonically decreases.\,\,The
frequency obtained from experiments in the argon matrix \cite{Buck}
equals 3574~\textrm{cm}$^{-1}$.\,\,Therefore, we believe that it is
possible to talk about a complete coincidence of calculated and
experimental results.\,\,From our viewpoint, this is a powerful
argument in favor that the frequency shift of valence vibrations has
the electrostatic origin.


\begin{table}[b]\label{tab6}
\noindent\caption{Frequencies of symmetric\\ valence vibrations of
hydrogen atoms\\ in the water molecule in vapor, water, and
ice}\vskip3mm\tabcolsep8.3pt

\noindent{\footnotesize\begin{tabular}{|l|c|c|c|}
  \hline%
 \multicolumn{1}{|c|}{\rule{0pt}{5mm}   }%
 & \multicolumn{1}{|c|}{$\nu _{\rm v},~\text{сm}^{ - 1}$ }
 & \multicolumn{1}{|c|}{$\nu _{\rm  w},~\text{сm}^{ - 1}$}
 & \multicolumn{1}{|c|}{$\nu _{\rm {Ice}},~\text{сm}^{ - 1}$}
  \\[2mm]
  \hline
\rule{0pt}{5mm}Experim.~\cite{Eizenberg} & 3657 & 3490 & 3200 \\

  Experim.~\cite{Zhu} & ~~3656.7 & 3280 &  \\

  MST-FP~\cite{Zhu} & 3656 & 3251 & \\

  SPC-FP~\cite{Zhu} & & 3875 & \\[2mm]
  \hline
\end{tabular}}
\end{table}

It is very important that the explanation of the frequency shifts
different by magnitude is principally based on the superposition
principle, the application of which to sharply directed and
saturated irreducible hydrogen bonds is impossible.\,\,In order to
make the substantiation of this fact more complete, we intend to
consider the frequency shifts of valence vibrations in ice and
liquid water elsewhere.

Not less demonstrative is the circumstance that one should expect a
positive sign of the frequency shift in rarefied vapor, which
directly follows from the behavior of
$\Delta\tilde{\mathrm{K}}_{\mathrm{rr}}^{{(1)}}/\tilde{\mathrm{K}}_{\mathrm{rr}}^{{(1)}}$
in Fig.~6.\,\,This fact is also supported qualitatively by the
experimental data on IR absorption in rather rarefied water vapor
\cite{Jin}.

\section{Influence of Neighbor Molecules\\ on the Dipole Moment of a Water
Molecule}

The dipole moment of an isolated water molecule is determined as a sum of two
oppositely directed vectors of dipole moments, $\boldsymbol{\mu}%
={\boldsymbol{\mu}}_{\mathrm{H}}+{\boldsymbol{\mu}}_{\mathrm{O}}$.\,\,The
dipole moment ${\boldsymbol{\mu}}_{\mathrm{H}}$ is determined by the
spatial distribution of the centers of oxygen's negative charge and
hydrogens'
positive charges, ${\boldsymbol{\mu}}_{\mathrm{H}}=q_{\mathrm{H}}%
(\mathbf{r}_{1}+\mathbf{r}_{2})$.\,\,The absolute value of dipole
moment
${\boldsymbol{\mu}}_{\mathrm{H}}$ equals $\mu_{\mathrm{H}}=$ $=2q_{\mathrm{H}%
}r_{\mathrm{OH}}\cos\left(  \theta/2\right)  =5{.}6281$~D.\,\,The
dipole momentum of the oxygen atom ${\boldsymbol{\mu}}_{\mathrm{O}}$
emerges owing to the polarization of the oxygen anion's electron
shell in the electric field created by hydrogen atoms in the water
molecule.\,\,According to work \cite{Stillinger}, it equals
\[
{\boldsymbol {\mu }}_{\rm O}  = - \alpha q_{\rm H} \left(\!
{\frac{{{\bf{r}}_1 }}{{r_1^3 }}\left[ {1 - K(r_1 )} \right] +
\frac{{{\bf{r}}_2 }}{{r_2^3 }}\left[ {1 - K(r_2 )} \right]}\!
\right)\!.
\]
It is easy to make sure that
$\mu_{\mathrm{O}}=-3.7752$~D.\,\,To\-gether with
${\boldsymbol{\mu}}_{\mathrm{H}}$, the following value is obtained
for the
absolute value of dipole moment $\boldsymbol{\mu}$: $\mu=\mu_{\mathrm{H}%
}+\mu_{\mathrm{O}}=$ $=1.8528$~D.\,\,It completely agrees with the
absolute value of dipole moment in an isolated water molecule.

\begin{figure}
\vskip1mm
\includegraphics[width=\column]{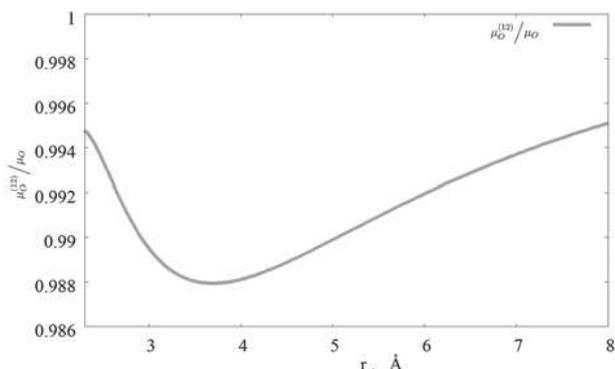}
\vskip-3mm\caption{Dependence of the ratio
{$\mu_{\mathrm{O}}^{(12)}/\mu_{\mathrm{O}}$} on the distance between
oxygen atoms in two water molecules
}\label{f35}
\end{figure}

The variation of the dipole moment under the action of a neighbor
molecule is one of the simplest manifestations of many-particle
effects in the system.\,\,To estimate the influence of the second
molecule, let us calculate the ratio
$\mu_{\mathrm{O}}^{(12)}/\mu_{\mathrm{O}}$ between the dipole
moments of the oxygen atom in the pair approximation
($\mu_{\mathrm{O}}^{(12)}$, see work \cite{Zhyganiuk}) and in the
isolated water molecule ($\mu_{\mathrm{O}}$).\,\,The calculated
dependence of $\mu_{\mathrm{O}}^{(12)}/\mu_{\mathrm{O}}$ on the
distance between the oxygen atoms in two neighbor molecules is
depicted in Fig.~7.

One can see from Fig.~7 that the variation of the dipole moment of
the oxygen atom under the influence of the electric field created by
the neighbor molecule does not exceed 1.5\%.\,\,The same can also be
said about the component ${\boldsymbol{\mu}}_{\mathrm{H}}$.\,\,From
whence, it follows that the effects of electron shell overlapping,
which are responsible for the change of the dipole moment of a water
molecule, are insignificant.\,\,This result completely agrees with
the conclusions of works \cite{Dolgushin76, Fulton, Barnes,
BerendsenVelde}.\,\,Consequently, this means that the irreducible
components of the interaction between molecules, which have to be
regarded as hydrogen bonds, are much smaller in comparison with the
energy of electrostatic interaction between molecules.

\section{Arguments in Favor\\ of the Existence of a Hydrogen Bond }

In the previous sections, we presented the facts, whose explanation
does not need the hypothesis about the existence of hydrogen bonds
in water and other classical liquids according to L.A.~Bulavin's
classification \cite{Bulavin}.\,\,However, in this section, we
describe a phenomenon, which cannot be explained without attracting
the hydrogen bond concept.\,\,This is the temperature dependence of
the water heat capacity.\,\,For convenience, this parameter will be
reckoned in the dimensionless units $i_{Q}={2C_{V}}/\left(
k_{\mathrm{B}}N_{\mathrm{A}}\right)  $, where ${C_{V}}$ is the heat
capacity of a gram molecule, and $N_{\mathrm{A}}$ Avogadro's
constant.\,\,These dimensionless units for the heat capacity will be
called the number of thermal degrees of freedom.

The latter differs from the standard number of degrees of freedom in
that the number of vibrational degrees of freedom in it
doubles.\,\,Argon can serve as the simplest example here.\,\,In
rarefied vapor, an argon atom is described by three independent
coordinates, which give its spatial position.\,\,The corresponding
value of $i_{Q}$ also equals 3.\,\,In the crystalline state of
argon, the number
of standard degrees of freedom for its atoms also equals 3.\,\,However, $i_{Q}%
=6$, because each degree of freedom corresponds to the vibrational motion.

The number $i_{Q}$ of thermal degrees of freedom per water molecule consists
of three components,
\[
i_{Q}=i_{Q}^{(\mathrm{tr})}+i_{Q}^{(\mathrm{or})}+i_{Q}^{(\upsilon)},
\]
which correspond to the translational motion of the molecules, their rotation,
and probable vibrations of irreducible hydrogen bonds, which are formed in all
water phases \cite{Benjamin, Goldman}.

From Fig.~8, one can see that the maximum value of heat capacity
equals 6 for Ar in the liquid phase, 12 for hydrogen sulfide, and
reaches 20 and even more for water.\,\,The maximum value corresponds
to a situation where every of the ordinary degrees of freedom has a
vibrational character.

The heat capacity of liquid argon turns out some lower than in the
solid state; nevertheless, it is close to 6.\,\,In the case of
hydrogen sulfide, three orientational degrees of freedom have also
to be taken into account.\,\,If they have been vibrational, the
maximum $i_{Q}$-value would have been close to 12.\,\,Water would
also have had this value for the number of thermal degrees of
freedom per molecule if its molecules have not been bound with one
another by means of hydrogen bonds.\,\,However, actually, as one can
see, the number $i_{Q}$ for water exceeds this value
approxima\-tely~by 6.

From the physical viewpoint, this difference has a natural
explanation: there are weak hydrogen bonds between molecules, which
practically do not interfere with the rotational motion of the
molecules, but make additional contributions to the water heat
capacity.\,\,These contributions result from two transverse and one
longitudinal vibrations of the hydrogen bond.\,\,Since every
vibration corresponds to two thermal degrees of freedom, the
excitation of all vibrations for only one hydrogen bond results in
the heat capacity growth by 6.

A more detailed analysis of the problem shows \cite{Lishchuk} that, in order to
attain a complete agreement with experimental data, it suffices to admit that
every molecule of liquid water forms 2.5 hydrogen bonds near the triple point
temperature and only one hydrogen bond in a vicinity of the critical point.
Those estimations are in quite satisfactory agreement with the results of
works \cite{Zakharchenko, Makhlaichuk2013}, as well as with the results of
computer calculations.

\section{Hydrogen Bond\\ from the Viewpoint\\ of the Chemical Bond Theory}

The hydrogen bond concept seems to appear for the first time in the
work by A.R.~Hantzsch in 1909 \cite{Hantzsch}.\,\,Using the water
molecule as an example, the sense of this new concept can be
interpreted in the following way.\,\,The hydrogen bond is a new type
of interaction between two water molecules.\,\,It acts along the
O--H--O line and is associated with the emergence of a specific
interaction between the indicated groups of molecules that arises at
certain distances between them.\,\,The introduced interaction is
much stronger than the van der Waals one, but, at the same time, is
much weaker than the covalent bond and the ionic interaction.\,\,In
work \cite{Latimer}, an attempt was made to identify this specific
interaction with the covalent bond between the hydrogen atom, which
can hold two electron pairs about itself, and two electronegative
atoms.\,\,L.~Pauling criticized this approach \cite{Pauling} and
presented arguments in favor of the ionic nature of a hydrogen bond.
Supposing that the hydrogen bond has a sharply directed character
and is saturated, i.e.\,\,it does not agree with the superposition
principle, L.~Pauling calculated the residual entropy of ice
($=k_{\mathrm{B}}\ln(3/2)$) and showed that this value agrees well
with experimental data (see also work \cite{LokotoshGorun}).\,\,This
result promoted the further propagation of the hydrogen bond concept
while describing the properties of ice, water, alcohols, and so
forth.


\begin{figure}[b]
\vskip-2mm
\includegraphics[width=\column]{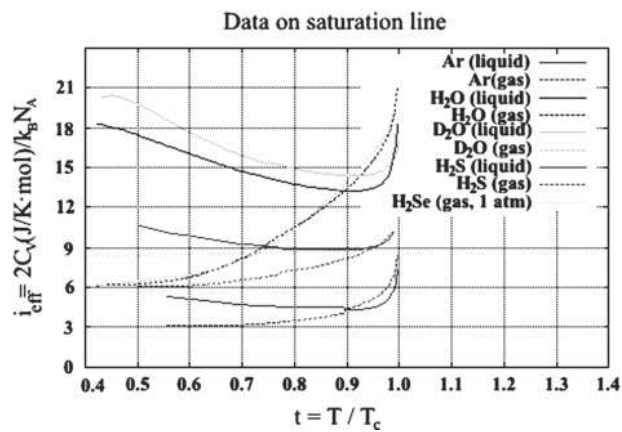}
\vskip-3mm\caption{Temperature dependences of {$i_{\mathrm{eff}}$ }for {$\mathrm{H}%
_{\mathrm{2}}\mathrm{O}$, $\mathrm{D}_{\mathrm{2}}\mathrm{O}$, $\mathrm{H}%
_{\mathrm{2}}\mathrm{S}$, and argon }systems in the liquid and vapor
states at the curves of their coexistence. Experimental data are
taken from work~{\cite{NIST}} }\label{f8}\vspace*{1.5mm}
\end{figure}

\section{Discussion of the Results Obtained}

\begin{figure}[b!]
\vskip-2mm
\includegraphics[width=\column]{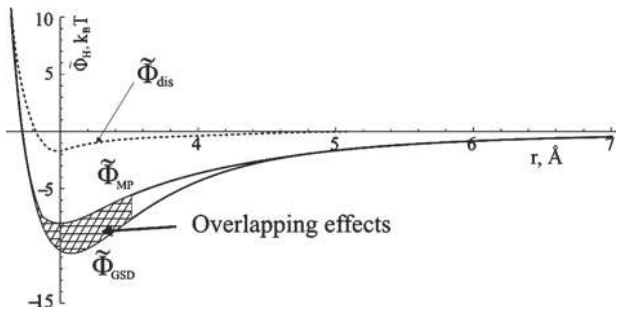}
\vskip-3mm\caption{ Partial contributions to the interaction
potential between water molecules in the linear dimer }\label{f47}
\end{figure}

\begin{figure}[b!]
\vskip1mm
\includegraphics[width=\column]{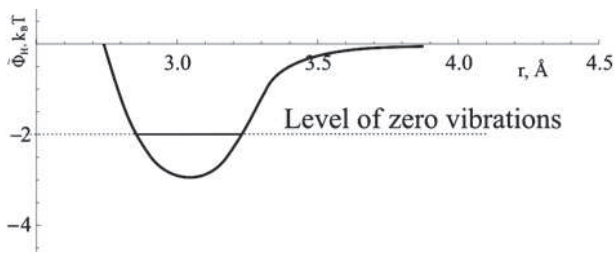}
\vskip-3mm\caption{ Hydrogen bond potential in the linear dimer of
water molecules }\label{f48}\vspace*{1.5mm}
\end{figure}

Let us summarize the results presented above, being based on the
general ideas concerning the structure of intermolecular potentials
in classical systems.\,\,We proceed from the fact that the simplest
structure of the interaction between molecules is inherent to atomic
systems of the argon type.\,\,The corresponding potential of the
intermolecular interaction $\Phi(r)$ is a sum of an attractive
component $\Phi_{\mathrm{dis}}(r)$, which is associated with
dispersion forces, and a component $\Phi_{\mathrm{rep}}(r)$
describing the repulsion:
\begin{equation}
\Phi(r)=\Phi_{\mathrm{dis}}(r)+\Phi_{\mathrm{rep}}(r).\label{eq42}%
\end{equation}
Note that the known Lennard-Jones potential has just this structure.

The systems consisting of molecules like N$_{2}$ are characterized by the loss of
spherical symmetry, which is accompanied by the appearance of the appreciable
angular dependence in the particle-to-particle potential \cite{Krokston}:
\begin{equation}
\Phi(r)\rightarrow\Phi(r,\Omega)=\Phi_{\mathrm{dis}}(r,\Omega)+\Phi
_{\mathrm{rep}}(r,\Omega). \label{eq43}%
\end{equation}
This structure must also be inherent to the interaction potentials between
water and alcohol molecules if the distance between them considerably exceeds
the sum of their molecular radii.

However, as the molecules get closer, their electron shells overlap, and a
qualitatively new component of the intermolecular interaction, $\Phi_{\mathrm{H}%
}(r,\Omega)$, appears, which is conventionally called the hydrogen
bond energy.\,\,In this case,
\[
\Phi(r,\Omega)=\Phi_{\mathrm{dis}}(r,\Omega)+\Phi_{\mathrm{rep}}(r,\Omega)\,+
\]\vspace*{-7mm}
\begin{equation}
+\,\Phi_{E}(r,\Omega)+\Phi_{\mathrm{H}}(r,\Omega).\label{eq44}%
\end{equation}
At the same time, the hydrogen bond energy, as a rule (see work
\cite{Mahlaichuk}), is associated with the sum of two last terms,
\begin{equation}
E_{\mathrm{H}}(r,\Omega)=\Phi_{E}(r,\Omega)+\Phi_{\mathrm{H}}(r,\Omega
),\label{eq45}%
\end{equation}
and M.D.~Sokolov was the first who attracted attention to this fact for the
first time \cite{Sokolov}.

In order to estimate the irreducible contribution to the hydrogen
bond energy, the following approach was proposed in work
\cite{Mahlaichuk}.\,\,From the conventional viewpoint, the ground
state energy of a water dimer is determined by the \textquotedblleft
hydrogen bond\textquotedblright\ energy
$E_{\mathrm{H}}(r_{d},\Omega_{d})$, where the subscript $d$
indicates that the distance between the oxygen atoms of two water
molecules and the angle values correspond to the dimer
configuration.\,\,On the other hand, the properties of dimers are
described quite well, by using phenomenological intermolecular
potentials of the SPC~\cite{Berendsen}, SPC/E~\cite{BerendsenSPCE},
TIPS~\cite{Jorgensen2}, SD~\cite{Stillinger}, GSD~\cite{Zhyganiuk},
and so forth types.

The dimer configuration is determined from the condition of
interaction energy minimum.\,\,Proceeding from this fact and
expression (\ref{eq44}), the energy of an irreducible hydrogen bond
can be estimated with the help of the relation
\[
\Phi_{\mathrm{H}}(r,\Omega)=\Phi(r,\Omega)\,-
\]\vspace*{-7mm}
\begin{equation}
-\,\left[  {\Phi_{\mathrm{dis}}(r,\Omega)+\Phi_{\mathrm{rep}}(r,\Omega
)+\Phi_{M}(r,\Omega)}\right]  \!,\label{eq46}%
\end{equation}
where $\Phi_{M}(r,\Omega)$ is the multipole approximation to the
electrostatic interaction energy.\,\,The multipole moments of a
water molecule are assumed to be determined independently of
effective charges governing the behavior of the phenomenological
potential $\Phi(r,\Omega)$.\,\,The magnitudes of effective charges,
as well as some other parameters of phenomenological potentials, are
determined by fitting the model to dimer parameters, which are
determined experimentally or within quantum-chemical
methods.\,\,Hence, it is the difference on the right-hand side of
Eq.~(\ref{eq46}) that describes the component emerging owing to the
overlapping of electron shells in water molecules.\,\,In work
\cite{Mahlaichuk}, the intermolecular potential $\Phi(r,\Omega)$ was
simulated by the generalized Stillinger--David potential
$\Phi_{\mathrm{GSD}}(r)$, so that
\[
\Phi_{\mathrm{H}}(r,\Omega)=\Phi_{\mathrm{GSD}}(r,\Omega)\,-
\]\vspace*{-7mm}
\begin{equation}
-\left[  {\Phi_{\mathrm{rep}}(r,\Omega)+\Phi_{\mathrm{dis}}(r,\Omega)+\Phi
_{M}(r,\Omega)}\right]  \!.\label{eq47}%
\end{equation}
The general behavior of the potentials $\Phi_{\mathrm{GSD}}(r,\Omega)$,
$\Phi_{M}(r,\Omega)$, and $\Phi_{\mathrm{dis}}(r,\Omega)$ for the molecular
orientation typical of the dimer are exhibited in Fig.~9.

The hydrogen bond potential for the same configuration of water
molecules is shown in Fig.~10.\,\,This is a short-range potential
that emerges owing to the overlapping of electron shells and has a
quantum-mechanical origin.\,\,It is this potential that should be
interpreted as the hydrogen bond potential in water.\,\,Its depth,
by order of magnitude, is identical to the depth of the dispersion
force interaction potential between water molecules, but it is
substantially smaller by magnitude than the multipole interaction
potential.\,\,As a result, the contribution made by hydrogen bonds
to the thermodynamic potentials of water can be taken into account
in the framework of the thermodynamic perturbation theory.\,\,At the
qualitative level, this circumstance completely agrees with the
similarity of the thermodynamic functions of water and argon at
their coexistence curves.

The presented estimations of the magnitudes of the electrostatic
interaction and hydrogen bond contributions to the energy of
intermolecular interaction were confirmed many times in works
\cite{Zhyganiuk, Dolgushin76, Fulton, Barnes, BerendsenVelde}.\,\,In
particular, it was shown in work \cite{Fulton} that, by order of
magnitude,
\[
\frac{\left\vert U_{E}(q_{1},q_{2})\right\vert }{\left\vert U_{\mathrm{H}%
}(q_{1},q_{2})\right\vert }\thicksim\frac{10}{2}.
\]
This means that the influence of exactly hydrogen bonds is taken into account in
a natural way by the thermodynamic perturbation theory \cite{Lishchuk,
Lishchuk10}.

While analyzing the thermodynamic properties of liquids and
solutions with the use of statistical theory, it is necessary to
take into consideration the fact that molecules permanently rotate
and, at the same time, hinder their rotational motion.\,\,The
characteristic period of the thermal rotational motion of molecules
turns out much shorter than the characteristic time of changing the
configurations formed by the translational degrees of freedom.
Therefore, the thermodynamic properties of liquids are mainly
governed by the potentials $U_{A}(r_{12})$ averaged over all angular
variables (here, $r_{12}$ is the distance between the centers of
mass of molecules).\,\,In works \cite{Lishchuk, Lishchuk10}, it was
demonstrated that such an averaged potential has a structure of the
Sutherland potential,\vspace*{-1mm}
\begin{equation}
U_{A}(r_{12})\Rightarrow\left\{  \!\!{%
\begin{array}
[c]{ll}%
{\infty,} & {r_{12}<r_{0}}\\
{U_{a}(r_{12}),} & {r_{12}>r_{0}},
\end{array}
}\right.  \label{eq48}%
\end{equation}
where $U_{a}(r)$ is the attraction potential, which decreases at
large enough distances following the law $1/r_{12}^{6}$.\,\,With the
same accuracy, the averaged potential can be approximated by the
Lennard-Jones potential \cite{Lishchuk, Lishchuk10}:\vspace*{-1mm}
\begin{equation}\label{eq49}
\begin{array}{l}
\displaystyle {U_A (r_{12} ) \Rightarrow U_{\rm LJ} (r_{12} ),}  \\[1mm]
\displaystyle {U_{\rm LJ} (r_{12} ) =  - 4\varepsilon \bigg[
{\bigg(\! {\frac{\strut \sigma } { {r_{12} }}} \!\bigg)^{\!\!12} -
\bigg(\! {\frac{\strut \sigma }{{r_{12} }}} \!\bigg)^{\!\!6} }
\bigg]}\!.
\end{array}
\end{equation}

The thermodynamic properties of liquids consisting of anisotropic
molecules, owing to the rotational motion of those molecules, are
similar to the corresponding properties of atomic liquids of the
argon type \cite{Lishchuk, Lishchuk10}.\,\,Small differences between
those properties stem from weak angular correlations, which can be
taken into account with the help of perturbation theory.\,\,The most
noticeable angular correlations are observed in the temperature
interval of the supercooled liquid state and at the formation of an
instant local structure in liquids.\,\,This circumstance is a
precondition for the emergence of special points in aqueous alcohol
solutions.

From the qualitative viewpoint, such a change of priorities is not justified,
because the analytic continuation of the components $\Phi_{\mathrm{dis}%
}(r,\Omega)$ and $\Phi_{M}(r,\Omega)$ into the region, where
electron shells overlap is not accompanied by the appearance of
effects that would violate the requirement of continuity in the
potential behavior.\,\,For this reason, it is desirable that the
hydrogen bond potential should be defined in another
way.\,\,According to the continuity requirement, the hydrogen bond
potential will be defined, by using the formula\vspace*{-1mm}
\[
\Phi(r,\Omega)=\Phi_{\mathrm{dis}}(r,\Omega)\,+
\]\vspace*{-9mm}
\begin{equation}
+\Phi_{r}(r,\Omega)+\Phi_{M}(r,\Omega)+\Phi_{\mathrm{H}}(r,\Omega
).\label{eq50}%
\end{equation}
In the region where the electron shells overlap, the functions $\Phi
_{\mathrm{dis}}(r,\Omega)$ and $\Phi_{M}(r,\Omega)$ should be substituted by
their corresponding continuations from the region, where their application does
not invoke doubts.

The formation of dimers and multimers of higher orders in vaporous
and liquid water is one of the most characteristic manifestations of
hydrogen bonds, which are responsible for the specificity of the
interaction between molecules.\,\,In other words, the study of the
properties of dimers in water provides us with direct information
concerning the properties of hydrogen bonds.\,\,This circumstance
gives us an exact instruction on how one must approach the research
of the properties of the interaction between molecules in water and
the hydrogen bond formation itself.\,\,The main stages of such an
approach are as follows.\,\,First, to describe the energy of
interaction between two water molecules, the most suitable
phenomenological model potential is selected among those that
describe the ground-state energy of a dimer the most
successfully.\,\,At the second stage, the obtained dependence of the
interaction energy of water molecules on the distance between them
is compared with the interaction energy determined from the
asymptotic multipole series expansion.\,\,Finally, in order to
determine the dependence of the hydrogen bond energy on the distance
between molecules, the difference between the energies of the model
potential and the sum of the dispersion and multipole components is
calculated.\,\,This difference is expected to be different from zero
only within a certain vicinity of the equilibrium distance between
water molecules \mbox{in the dimer.}\looseness=2

In order to describe the interaction between molecules in a dimer,
we use the generalized Stil\-lin\-ger--Da\-vid potential proposed in
work \cite{Zhyganiuk}.\,\,It is a soft potential, whose parameters
can be changed due to the interaction with neighbor
molecules.\,\,This is a very important circumstance, which cannot be
taken into account in the majority of phenomenological model
potentials \cite{Jorgensen2, Poltev, Matsuoka, Hirschfelder,
Eizenberg, Rieth}.\,\,Unlike the original Stil\-lin\-ger--Da\-vid
potential \cite{Stillinger}, its generalized variant (GSD) more
adequately involves the behavior of screening functions, which
describe the effects of electron shell overlapping.\,\,In addition,
the Stil\-lin\-ger--Da\-vid potential was corrected with respect to
it asymptotic behavior at large enough distances between molecules,
when it should be determined by the dipole-dipole interaction.

One can see that the depth of the irreducible component $\Phi_{\mathrm{H}%
}(r,\Omega)$ in the interaction potential between water molecules
resulting from the overlapping of their electron shells does not
exceed $(2$$\div$$ 3)k_{\mathrm{B}}T_{m}$.\,\,By order of magnitude,
it is close to the dispersion component
$\Phi_{\mathrm{dis}}(r,\Omega),$ but is substantially smaller in
comparison with the multipole interaction potential
$\Phi_{M}(r,\Omega)\approx (7$$\div$$8)k_{\mathrm{B}}T_{m}$.

Actually, the insignificant depth of the potential well formed by
the hydrogen bond results in that the contributions of the
interaction potential component to the thermodynamic potentials and
the kinetic coefficients can be taken into account in the framework
of perturbation theory.\,\,In addition, the temperature behavior of
main thermodynamic parameters of water such as the fractional
molecular volume, evaporation heat, and others, has an argon-like
character with a quite satisfactory accuracy.\,\,These conclusions
are completely confirmed by the results of works \cite{Lishchuk,
Fisenko, Lishchuk10}.

The hydrogen bond potential plotted in Fig.~10 corresponds to the
relative orientation of water molecules in the equilibrium dimer
configuration only.\,\,In principle, there are no complications for
the construction of the potential $\Phi_{\mathrm{H}}(r,\Omega)$ at
all other relative orientations of water molecules, since the
angular dependences for the potentials $\Phi
_{\mathrm{GSD}}(r,\Omega)$ and $\Phi_{M}(r,\Omega)$ are known for
arbitrary angles.

The conclusion about a weak deformation of electron shells and, as a
consequence, the formation of weak irreducible hydrogen bonds is obviously
supported by the results of work \cite{Weiss}, in which a redistribution of
the electron density was analyzed, by using the methods of scanning tunnel microscopy.

It should be noted that the thermodynamic properties of water are
determined by the potentials averaged over the angles, which is a
consequence of the rotational motions of water molecules.\,\,Owing
to this averaging, as was shown in work \cite{Lishchuk10}, the
potential well depth of a hydrogen bond additionally decreases,
which gives rise to a correction of the argon-like dependences for
the thermodynamic parameters, whose relative magnitude does not
exceed 5\% \cite{Lishchuk, Lishchuk10}.\,\,At the same time, the
hydrogen bonds manifest themselves directly in the water heat
capacity \cite{Lishchuk}.\,\,Another important circumstance falling
beyond the scope of our consideration is the adequate account for
the environment influence on the character of the hydrogen bond
potential.\,\,We are planning to consider this issue in detail
\mbox{elsewhere.}\looseness=1

\vskip3mm

\textit{This work would be impossible without regular consultations
over the years with Profs.\,\,T.V.~Lokotosh, G.G.~Malenkov,
Yu.I.~Naberukhin, G.O.~Puchkovska, and V.E.~Pogorelov.\,\,We are
also grateful to our coauthors P.V.~Makhlaichuk and S.V.~Lishchuk.
While carrying out this and other works in this direction, we felt
the constant support of Academician of the NAS of Ukraine
L.A.~Bulavin, for which we are sincerely thankful to him.}

\textit{The results of those works were reported at various seminars
and conferences.\,\,After the first reports were made, a careful
criticism of attendees was gradually changed in favor of our
viewpoint.}

\rezume{%
І.В.\,Жиганюк, М.П.\,Маломуж}{ФІЗИЧНА ПРИРОДА ВОДНЕВОГО ЗВ'ЯЗКУ} {У
роботі досліджується фізична природа та коректність означення
водневих зв'язків. Аналізується, перш за все, вплив останніх на
поведінку термодинамічних, кінетичних та спектроскопічних
властивостей води. Показано, що сприйняття водневих зв'язків як
гостронаправлених та насичених містків, які виникають між молекулами
води, є несумісним з поведінкою специфічного об'єму та теплоти
випаровування, а також коефіцієнтів самодифузії та кінематичної
зсувної в'язкості. На додаток до цього показано, що зміна дипольного
моменту молекул води, а також зсув частоти валентних коливань
гідроксильної групи повністю пояснюються на основі уявлень про
електростатичну природу водневого зв'язку. Разом з тим, температурні
залежності теплоємності води та її пари чітко вказують на існування
слабких водневих зв'язків. Аналізуючи властивості димеру води,
показано, що внесок слабких водневих зв'язків у енергію основного
стану димеру є приблизно в 4--5 разів меншим у порівнянні з енергією
електростатичної взаємодії між молекулами води. Підсумовуючи
результати, робиться висновок, що таку саму природу мають водневі
зв'язки в усіх інших випадках, де вони виникають.}

\end{document}